\documentstyle[12pt,epsfig]{article}

\newlength{\dinwidth}
\newlength{\dinmargin}
\begin{document}
\def\bold#1{\setbox0=\hbox{$#1$}%
     \kern-.025em\copy0\kern-\wd0
     \kern.05em\copy0\kern-\wd0
     \kern-.025em\raise.0433em\box0 }
\def\slash#1{\setbox0=\hbox{$#1$}#1\hskip-\wd0\dimen0=5pt\advance
       \dimen0 by-\ht0\advance\dimen0 by\dp0\lower0.5\dimen0\hbox
         to\wd0{\hss\sl/\/\hss}}
\def\lq{\left [}
\def\rq{\right ]}
\def\II{{\cal I}}
\def\LL{{\cal L}}
\def\VV{{\cal V}}
\def\AA{{\cal A}}
\def\BB{{\cal B}}
\def\MM{{\cal M}}
\def\pv{\mbox{\bf p}}
\def\ovl{\overline}
\def\pr{^\prime}
\newcommand{\be}{\begin{equation}}
\newcommand{\ee}{\end{equation}}
\newcommand{\bea}{\begin{eqnarray}}
\newcommand{\eea}{\end{eqnarray}}
\newcommand{\ba}{\begin{array}}
\newcommand{\ea}{\end{array}}
\newcommand{\nn}{\nonumber}
\newcommand{\dd}{\displaystyle}
\newcommand{\bra}[1]{\left\langle #1 \right|}
\newcommand{\ket}[1]{\left| #1 \right\rangle}
\newcommand{\spur}[1]{\not\! #1 \,}
\newcommand{\nor}[1]{{}_\times^\times #1 {}_\times^\times}
\newcommand{\doo}[1]{\delta^{oo}_{{#1}}}
\newcommand{\dee}[1]{\delta^{ee}_{{#1}}}
\newcommand{\deo}[1]{\delta^{eo}_{{#1}}}
\newcommand{\doe}[1]{\delta^{oe}_{{#1}}}

\thispagestyle{empty}
\vspace*{1cm}
\rightline{Napoli DSF-T-58/96}
\vspace*{2cm}
\begin{center}
  \begin{LARGE}
  \begin{bf} 
${\bf {\cal W}_{k}}$ structure of generalized Frenkel-Kac construction for 
${\bf SU(2)}$-level ${\bf k}$ Kac-Moody algebra 
  \end{bf}
  \end{LARGE}

  \vspace{8mm}

  \begin{large}
 Vincenzo Marotta$^{\dag}$
  \end{large}
  \vspace{1cm}

\begin{it}
$^{\dag}$ 
Dipartimento di Scienze Fisiche  \\
 Universit\'a di Napoli ``Federico II'' \\ 
and \\ 
INFN, Sezione di Napoli
\end{it}
\end{center}
\begin{quotation}

\begin{center}
\begin{bf}
Abstract\\  
\end{bf}\end{center}
${\cal W}_{k}$ structure underlying the transverse realization of $SU(2)$ at 
level $k$ is analyzed. 
Extension of the equivalence existing between covariant and light-cone gauge 
realization of affine Kac-Moody algebra to ${\cal W}_{k}$ algebras is given.
Higher spin generators related to parafermions are extracted from the operator 
product algebra of the generators and are showed to be written in terms of 
only one free boson compactified on a circle. 
\noindent

\vspace*{0.5cm}



\vfill
{\small\bf
\begin{tabbing}
Postal address: Mostra d'Oltremare Pad.19-I-80125 Napoli, Italy \\ 
E:mail: \=  vincenzo.marotta@na.infn.it 
\end{tabbing}}
\end{quotation}

\newpage
\baselineskip=18pt
\setcounter{page}{2}

\section{Introduction}

At present there is a renewed interest in non standard realization of Kac-Moody 
(KM) algebras due to their relationships with new results in string theory and 
duality \cite{DVV}.

In this paper I discuss how some of these structures appear naturally in the 
projection of covariant realization of affine KM algebras and transverse one 
introduced in a my old paper in the relationship with the attempt to 
describe Lorentzian KM algebras \cite{VM}.

As it is well known \cite{FI} there are two way to give a realization of affine 
algebras to the level $k>1$ from the $k=1$ ones.

In the first case $k$ copies of representations are needed and the diagonal 
sub-algebra in their tensor product transforms as the same Kac-Moody algebra to 
the level $k$.

The second one makes use of the existence of a copy of the same algebra at the 
level $k$ embedded into the realization at $k=1$.
In fact, for any algebra, it is very easy to verify that the modes 
$J^{a}_{kn}$ realize a representation of the same algebra to the level $k$. 

While the first technique was extensively used in the past to study $k>1$ 
representations the second one has attracted the attention of physicists only 
recently in connection with the computing of roots multiplicity for 
hyperbolic Kac-Moody algebras \cite{B}, algebraic geometry \cite{N} and related 
subject as BPS states symmetry \cite{HM}.

This lack of interest seems to be originated from the non linearity of this 
embedding that makes a quite difficult task the analysis of this kind of coset 
to the respect of the linearity of the diagonal embedding due to the 
independence of the $k$ copies of the algebra.

Using the realization of \cite{VM} I show that the new interesting properties 
of this special coset originate from the possibility to realize parafermionic 
fields from the free bosons living on the compactified target-space for the 
level one. 

Therefore, no further extension of the target-space is needed in this case and 
no additional field must be defined to increase the level.

Extension to the full ${\cal W}_{k}$ parafermionic algebra can be done in 
particular for $SU(2)_k$ case  showing that only one boson Fock space is needed 
for this realization.
This reveals that the most natural setting for this realization should be the  
${\cal W}$-strings.

The paper is organized as follow: in Section 2 I introduce the covariant 
realization of affine KM algebras and its relationship with the transverse 
one that reduces to the standard Frenkel-Kac-Segal \cite{FKS} for level one. 
Then, the connection with the parafermions fields for $k>1$ and their 
realization in terms of the free bosons is given.

In Section 3 the extension to ${\cal W}_{k}$ algebra of the correspondence 
between covariant and light-cone realization is discussed and I show also how 
their operators can be extracted from the operator product algebra (OPA) of 
the currents at level $k$.  

Finally, Section 4 is devoted to discuss further relevant aspects and 
prospectives of these kind of coset, with a particular emphasis to the 
connection with Lorentzian algebras  and conclusions are given.

\bigskip 

\section{ Covariant construction of KM algebras }

In this section I review some aspects of the realization of higher level 
affine Kac-Moody algebras of ref.\cite{VM}. 

The Goddard-Olive construction of affine KM algebras \cite{GO2}can be 
interpreted as a vertex construction of a Lie algebra in a singular lattice 
obtained by adding a light-like direction to the Euclidean lattice defining 
the horizontal finite Lie algebra.
The outer derivation is consistent with the extension of the singular lattice 
to a Lorentzian one and with the interpretation of the affine algebra as a 
sub-algebra of a Lorentzian one.

Let me shortly recall the essential steps of the covariant construction. 
I introduce an infinite set of annihilation and creation operators 
$a^{\mu}_{n}$ $n\in Z$, satisfying the commutation relations:
\be
\left[ a^{\mu}_{n},a^{\nu}_{m}\right]=ng^{\mu\nu}\delta_{n+m,0}
\ee
with $g^{\mu\nu}$ a Minkowskian diagonal metric and 
$a^{\mu\dag}_{n}=a^{\mu}_{-n}$.

The momentum operator of string is $a^{\mu}_{0}=p^{\mu}$ and satisfies the 
commutation relations:
\be
\left[ q^{\mu},p^{\nu}\right]=ig^{\mu\nu}
\ee
with the position operator $q^{\mu}$.

Then I introduce the usual Fubini-Veneziano fields:
\be
 Q^{\mu}(z)=q^{\mu}-ip^{\mu}lnz+i\sum_{n\neq 0} \frac{a^{\mu}_{n}}{n} z^{-n}
\label {eq: 4} 
\ee
and their derivatives
\be
Q^{\mu(1)}(z)=i\frac{d}{dz}Q^{\mu}(z)=\sum_{n}a^{\mu}_{n} z^{-n-1} 
\label {eq: 5}
\ee
(I only consider holomorphic fields here).

If one consider the roots $r$ belonging  to a Lorentzian lattice, it can be 
decomposed as:
\be
r=\alpha+nK^{+}+mK^{-}
\ee
with $K^{\pm2}=0$, $K^{+}\cdot K^{-}=1$ and $\alpha$ belonging to $\Lambda$, the 
horizontal Euclidean lattice of a simply-laced Lie algebra.

Then the vertex operator associated to a root $r$ is:
\be
 U^{r}(z)=:e^{ir\cdot Q(z)}:
\ee
where the dots indicate the normal ordering, with the usual property:
\be
U^{r \dag}(z)=U^{-r}(\frac{1}{z^{*}})
\ee

The affine sub-algebra is spanned, for the real roots, by:
\be
A^{\alpha+nK^{+}}=\frac{c_{\alpha}}{2\pi i} \oint dz U^{\alpha+nK^{+}}(z)
\ee
$c_{\alpha}$ being a cocycle, and for the imaginary roots by:
\be
H^{i}_{nK^{+}}=\frac{1}{2\pi i}\oint dz :Q^{i(1)}(z)U^{nK^{+}}(z):
\ee
where the $i$ index is restricted to the Euclidean lattice. 

A quite general construction of cocycles has been done in \cite{GO} 
and I use those cocycles in the following construction.

The commutation relations are:
\bea
\left[ H^{i}_{nK^{+}},H^{j}_{mK^{+}}\right]&=& n \delta^{ij} \delta_{n+m,0} 
K^{+}\cdot p \\
\left[A^{\alpha+nK^{+}},A^{\beta+mK^{+}}\right] &=& 0 \hspace{6.1cm} \alpha \cdot 
\beta \geq 0 \\
\left[ A^{\alpha+nK^{+}},A^{\beta+mK^{+}}\right] &=& \epsilon (\alpha,\beta ) 
A^{\alpha+\beta+(n+m)K^{+}} \hspace{2.6cm} \alpha \cdot \beta =-1 \\
\left[ A^{\alpha+nK^{+}},A^{\beta+mK^{+}}\right] &=& \alpha \cdot H_{(n+m)K^{+}} 
+n\delta_{n+m,0} K^{+}\cdot p   \hspace{1cm} \alpha=-\beta  \label {eq: 14} \\
\left[ H^{i}_{nK{+}},A^{\alpha+mK^{+}}\right] &=& \alpha^{i} A^{\alpha+(n+m)
K^{+}}
\eea

One can define also a derivation (that does not belong to the affine KM 
algebra) by $D=-K^{-}\cdot p $ with the commutators:
\be
\left[ D,A^{\alpha+nK^{+}}\right]=-nA^{\alpha+nK^{+}} 
\ee
\be
\left[ D,H^{i}_{nK^{+}}\right]=-n H^{i}_{nK^{+}}
\ee

As  $K^{+}\cdot a_{n}$ commutes with any element of the algebra,
it is possible to take them to be a constant and particularly:
\be
 K^{+}\cdot p \rightarrow k \hspace{1 cm} K^{+}\cdot a_{n} \rightarrow 0 
\hspace{1cm} if \:\:\:\: n \neq 0 \label {eq: 19}
\ee
The level independence is now evident from the above construction and from 
eqs.(\ref {eq: 14}) and (\ref {eq: 19}). 

Let me emphasize that this property is a natural consequence 
of the extension of the Euclidean lattice to a Lorentzian one.

\subsection{ Relationship with parafermions}

The choice of eq.(\ref {eq: 19}) corresponds to a transformation from the 
covariant gauge to the transverse one, so I will examine this correspondence.

In this transformation the vertex operator $U^{nK^{+}}(z)$ is reduced to:
\be
U^{nK^{+}}(z)= e^{inK^{+}\cdot q}z^{nK^{+}\cdot p} 
\ee
where the phase $e^{inK^{+}\cdot q}$ is irrelevant in this context.

The other operators become:
\be
 U^{\alpha+nK^{+}}(z) \rightarrow z^{nk}:e^{i\alpha\cdot Q(z)}:, ~~~~~ 
 :Q^{i(1)}(z)U^{nK^{+}}(z):\rightarrow z^{nk}Q^{i(1)}(z) 
\ee
with modes:
\be
 A^{\alpha}_{n}=\frac{c_{\alpha}}{2 \pi i} \oint dz z^{nk} U^{\alpha}(z), ~~~~~ 
 H^{i}_{n}=\frac{1}{2\pi i} \oint dz z^{nk} Q^{i(1)}(z) 
\ee

If $ k=1$  this is the Frenkel-Kac-Segal \cite{FKS} construction, but if $k>1$ 
it looks as a new different construction. 

Now it is possible to discuss the connection of this construction with the 
parafermionic one \cite{GQ}.

For $k>1$ the Fock space $F^{k}$ of Heisenberg sub-algebra is build up by the 
set of $ H^{i}_{n} $ operators, which is the subset of the creation 
and annihilation operators of the whole Fock space $F $, $a^{\mu}_{m}$, with 
$\mu = i $ and $ m = nk $ with $ k $ fixed. 

Then I decompose the F space in $ F=F^{k}\otimes \Omega^{k}$, where 
$\Omega^{k}$ is the vector space of vacuum vectors for the Heisenberg 
sub-algebra. 

The Hilbert space where the operators of the  eq.(\ref {eq: 4}) act is:
\be
H=F\otimes \Lambda^{*} \label{eq: FOCK}
\ee
($\Lambda^{*}$ is the dual of $\Lambda$ lattice )

Let me now define the fields on the $F^{k}\otimes \Lambda^{*}$ space:
\be
 H^{i}(z^k)=\sum_{n} a^{i}_{nk} z^{-nk-1} 
\ee
(for $k=1$ from eq. (\ref {eq: 5}) $H^{i}(z^k)=Q^{i(1)}(z)$  ) with:
\be
 H^{i}(z^k) H^{j}(\xi^k)=: H^{i}(z^k) H^{j}(\xi^k) : + k \delta_{ij} 
\frac{z^{k-1} \xi^{k-1}}{(z^{k}-\xi^{k})^{2}} 
\ee
and:
\be
X^{i}(z^k)=q^{i}-ip^{i}lnz + i\sum_{n \neq 0}\frac{a^{i}_{n k}}{n k} z^{-n k} 
\label {eq: 27} 
\ee
One can also define a vertex operator on $F^{k}\otimes \Lambda^{*}$:
\be
{\cal U}^{\alpha}(z^k)=z^{1-\frac{1}{k}}:e^{i\alpha\cdot X(z^k)} : 
\label {eq: 100}
\ee
which satisfies the relation:
\be
 {\cal U}^{\alpha}(z^k) {\cal U}^{\beta}(\xi^k)=: {\cal U}^{\alpha}(z^k) 
{\cal U}^{\beta}(\xi^k):
(z^{k}-\xi^{k})^{\frac{\alpha \cdot \beta}{k}} 
\ee
 Moreover, the fields on $\Omega^{k}$ are defined by
\be
\psi^{\alpha}(z^k)=z^{\frac{1}{k}-1}: e^{i\alpha\cdot (Q(z)-X(z^k))} : 
\label {eq: 30} 
\ee
and satisfy
\be
 \psi ^{\alpha}(z^k) \psi ^{\beta}(\xi^k )=
:\psi ^{\alpha}(z^k) \psi^ {\beta}(\xi^k ) :\frac{(z-\xi)^{\alpha
\cdot \beta}}{(z^{k}-\xi^{k})^ {\frac{\alpha \cdot \beta}{k}}} \label {eq: 31}
\ee

These fields live on the $k$-sheeted complex plane which is the image space of 
the conformal transformation $z\rightarrow z^{k}$, thus, each value of the 
variable $z^{k}$ corresponds to $k$ different points on the plane related by 
a discrete transformation. They can be interpreted as the parafermionic fields 
for the $k$ level KM algebra.

I define an isomorphism between single-value fields on the k-sheeted 
complex plane and multi-values fields on the one-sheeted plane 
by means of the following identifications:
\be
a^{i}_{nk+l} \longrightarrow \sqrt{k}a^{i}_{n+l/k} \hspace{1cm} q^{i} 
\longrightarrow \frac{1}{\sqrt{k}}q^{i} \label {eq: 32}
\ee

This isomorphism acts on the target-space as a duality transformation 
$\alpha^2\rightarrow \alpha^{2}/k$ that reduces the compactification torus 
radius, therefore one can use it to move on between points of enhanced 
symmetry of the $d$ dimensional target-space of a string theory.

All the physical fields obtained with this isomorphism are single-valued by 
construction.

The fields of eq.(\ref {eq: 27}) become:
\be
X^{i}(z)= q^{i}-ip^{i}lnz +i\sum_{n\neq 0}\frac{a^{i}_{n}}{n}z^{-n} 
\label {eq: 33}
\ee
and the vertex ${\cal U}^{\alpha}(z) $ of the eq.(\ref {eq: 100}) becomes:
\be
{\cal U}^{\alpha}(z)=: e^{\frac{i\alpha\cdot X(z)}{\sqrt{k}}} : \label {eq: 34}
\ee

The above equations define the quantities appearing in the Gepner \cite{G}
construction in terms of these used in the Lorentzian lattice approach.

In fact the currents $ h^{i}(z)$ and $ \chi_{\alpha}(z) $ given by eq.(2) of 
ref. \cite{G} can be written as
\bea
h^{i}(z)&=&i\frac{d}{dz}X^{i}(z)=\sqrt{k} \sum_{n} a^{i}_{n} z^{-n-1} \label {eq: 35}  
\nn \\ 
\chi_{\alpha}(z)&=&c_{\alpha}\psi^{\alpha}(z){\cal U}^{\alpha}(z) 
\label {eq: 36}
\eea
and they satisfy the commutation relations of the eqs.(4) of the same reference.

By the definition eq.(\ref {eq: 30}) and by eq.(\ref {eq: 31}) 
using eqs.(\ref {eq: 32})$\div$(\ref {eq: 36}) we obtain the parafermionics 
relations:
\be
\psi ^{\alpha}(z) \psi ^{\beta}(\xi )= : \psi ^{\alpha}(z) \psi ^{\beta}(\xi 
) :
\prod^{k-1}_{p=1}
(z^{\frac{1}{k}}- \epsilon ^{p} \xi ^{\frac{1}{k}})^{-\alpha 
\cdot \beta}(z-\xi)^{\alpha \cdot \beta (1-\frac{1}{k})} \label {eq: PK} 
\ee
where \( \epsilon = e^{\frac{2\pi i}{k}} \).

For $\alpha \cdot \beta=-1 $  these 
relations give rise to the following OPE:
\be
\psi^{\alpha}(z) \psi^{\beta}(\xi)= k (z-\xi)^{-1+\frac{1}{k}} 
\left [\psi^{\alpha+\beta}(\xi) + {\cal O}(z-\xi)\right] \label {eq: 39}
\ee

and for  $ \alpha =- \beta $
\be
\psi^{\alpha}(z) \psi^{- \alpha}(\xi)= k^{2} (z-\xi)^{-2+\frac{2}{k}}
\left [1 + \frac{2 \Delta_\psi}{c_\psi}T_\psi(\xi)(z-\xi)^2+
{\cal O}(z-\xi)^3\right ] \label {eq: 38} 
\ee
where $T_\psi(\xi)$ is the stress-tensor, $c_\psi$ the central charge and 
$\Delta_\psi$ the conformal weight of parafermions.

Therefore by comparing eqs.(\ref {eq: 38}) and (\ref {eq: 39}) with eqs.(6) of 
ref. \cite{G} we can state that the fields $\psi_{\alpha}(z)$, given 
by eqs.(\ref {eq: 30}) and (\ref {eq: 31}), after a conformal transformation 
and the use of the isomorphism of eq.(\ref {eq: 32}) are proportional to the 
parafermions introduced in \cite{G}.

One can also decompose $\psi^{\alpha}(z) $ in a sum of $k$ parts with definite 
boundary conditions:
\be
\psi^{\alpha}(z)=\sqrt{k} \sum_{\lambda = 1}^{k} \psi^{\alpha}_{\lambda}(z)
\label {eq: BOUND} 
\ee
where $\psi^{\alpha}_{\lambda}(e^{2\pi i} z)=\epsilon^{\lambda}
\psi^{\alpha}_{\lambda}(z) $.
 
In this case the new modes become:
\be
A^{\alpha}_{n}=\frac{c_{\alpha} \sqrt{k}}{2 \pi i} \oint dz z^{n} {\cal U}^{\alpha}(z)
\psi^{\alpha}_{\lambda}(z)
\ee
where the boundary conditions for parafermionic fields are selected by the 
relation:
\be
\alpha \cdot p+\lambda = 0\:\: (mod \:\: k) \label {eq: ZK}
\ee
which is imposed in order to have a single-valued integrand, thus one must 
take a reduction in the $ \Omega ^{k} $ space that realizes the discrete 
symmetry derived from the charges in the coset $\Lambda ^{*}/k\Lambda $.

Let me emphasize once more that the parafermions appear naturally in this 
procedure and they are built on the bosonic space by means of bosonic fields 
through eq.(\ref {eq: 30}), so one can consider the present procedure to obtain 
parafermions as a generalized bosonization procedure.

Moreover the discrete symmetry acts only on these fields while the radius of 
target-space is only rescaled by the $k$ value.

\bigskip 

\section{Realization of ${\bf {\cal W}_{k}}$ algebras }

As it is well known to a parafermion system it is possible to associate a  
higher conformal spin extension of Virasoro algebra \cite{BBSS,ZF,FL}. 

I show that this can be done also for this realization by means of $k-1$ 
independents twisted bosonic fields $\phi^{l}(z)$.

The analysis is restricted only to the SU(2) affine algebra to level $k$, 
where only one Fubini field is needed and the discrete symmetry is $Z_{k}$.

In this case one can decompose the field $ Q(z)-X(z)$ into $k$ components:
\be
\phi^{l}(z^k) = Q(\epsilon^{l}z)-X(z^k) ~~~~\forall ~~~l=1,...,k  
\label {eq: TWIST}
\ee
with the constraint $\sum_{i=0}^{k}\phi^{l}(z^k)=0$, then the parafermions can 
be expressed by the following expression 
\be
\psi^{\pm}(z^k)=z^{\frac{1}{k}-1}
: e^{\pm i \sqrt{2}\phi^{1}(z^k)}: \label {eq: 43}
\ee
in terms of the $k-1$ independent components of eq:(\ref {eq: BOUND}) 

The operators:
\bea
H_{n}&=&\frac{1}{2\pi i} \oint dz z^{nk} H(z^k) \\ 
A^{\pm}_{n}&=&\frac{c_{\pm \sqrt{2}}}{2\pi i} \oint dz z^{nk} 
{\cal U}^{\pm}(z^k)\psi^{\pm}_{\lambda}(z^k)  \label{eq: 46}
\eea
are a realization of $SU(2)_{k}$ Kac-Moody algebra.

By means of the isomorphism defined by eq.(\ref {eq: 32}) one obtains an 
equivalent construction that is directly related to the standard realization 
\cite{G}, defining the free massless Bose chiral field $ X(z) $ 
eq.(\ref {eq: 33}) and the currents:
\bea
h(z)&=&\sqrt{k}i\frac{d}{dz} X(z)  \\
\chi^{\alpha}(z)&=&c_{\alpha}\sqrt{k} : e^{i \frac{\alpha X(z)}{\sqrt{k}}}:
\psi^{\alpha}_{\lambda}(z)
\eea
where $\alpha=\pm\sqrt{2}$ and $\psi_{\lambda}$ is in the sector satisfying 
the relation of eq:(\ref {eq: ZK}), obtaining the usual theory with a 
stress-tensor generating a Virasoro algebra with central charge  
$c=3k/(k+2)$ corresponding to level $k$  SU(2) affine algebra \cite{G,DT}.

Of course one of the most interesting application of the covariant vertex 
construction is the realization of Lorentzian algebras \cite{MS}, in fact, 
the  unified construction of arbitrary level representations of affine KM 
algebras appears quite naturally in this context, where the level $ k $ can 
be changed by the action of the pure Lorentzian generators, in complete 
analogy with the case of affine algebras where the weights of horizontal 
finite dimensional Lie sub-algebra are changed by the action of the affine 
generators.

Now I want to show how this realization naturally extends to ${\cal W}_{k}$ 
algebra of parafermions related to the coset $SU(2)_{k}/U(1)$ \cite{ZF}.

A standard construction of ${\cal W}_{k}$ with $Z_{k}$ symmetry make use of a 
vector consisting of $k-1$ scalar bosons $\varphi^{i}$ on the complex $z$ 
plane with untwisted boundary conditions that can be associated to the 
fundamental representation of $sl(k)$ algebra realized in $R^{k}$. 
These fields are related to a $k-1$ dimensional lattice $A_{k-1}$ and satisfy 
the relation $\sum_{i=1}^{k} \varphi^{i}(z) = 0 $ and normalization  
$\delta_{ij}-1/k$.

${\cal W}_{k}$ generators are defined expanding a generating function 
$R_{k}(z)$ obtained by means of quantum Miura transformation \cite{FL}
\be
R_{k}(z)=:\prod_{i=1}^{k}\left (\alpha_{0}\partial_{z} - 
i\partial_{z}\varphi^{i}(z)\right ): 
= -\sum_{n=0}^{k} {\cal W}^{n}(z) (\alpha_{0}\partial_{z})^{k-n} \label {eq: QM}
\ee

The lowest spin not trivial current is just energy-momentum tensor 
$T_{\varphi}(z)$:
\be
T_{\varphi}(z) = \sum_{i>j}^{k}
:\partial_{z} \varphi^{i}(z)\partial_{z} \varphi^{j}(z): + 
i \alpha_{0}\partial_{z}^{2}{\bf\rho}\cdot\varphi(z)
\ee
where ${\bf \rho}$ is the Weyl vector (half-sum of the fundamental weight 
vectors) for $A_{k-1}$ lattice with ${\bf\rho}^{2}=\frac{1}{12}k(k^{2}-1)$ and 
the central charge is given by:
\be
c_{\varphi} = (k-1)\left ( 1-\alpha_0^2 k(k+1)\right )
\ee

These algebra generators, in the Miura basis, are not primary nor 
quasi-primary, in general. 
It is possible to obtain a quasi-primary basis by a deformation of 
${\cal W}_k$ fields that involves only coefficients algebraic in $\alpha_0$ 
and, then, it does not modify algebraic structure.
Projection on primary basis is more difficult for the lack of a general 
algorithm. 
For instance, in the $n=3$ case this deformation is:
\be
{\cal W}^{3}(z^k)\rightarrow {\cal W}^{3}(z^k) - 
\frac{(k-1)}{2}\alpha_0 \partial_z{\cal W}^{2}(z^k)
\ee

Highest weight states of ${\cal W}_k$ algebra are created from vacuum $|0>$ 
by applying the vertex $:e^{i{\bf\lambda}\cdot\varphi (z)}:$ with $\lambda$ a 
highest weight of $sl(k)$.
 
Their conformal weights are:
\be 
\Delta_{\varphi}({\bf\lambda}) = \frac{1}{2}({\bf\lambda}\cdot ({\bf\lambda} 
-2\alpha_0 {\bf\rho})) 
\ee

It is well known that a conformal theory with the above central charge has a 
set of null fields depending on two vectors of parameters ${\bf r}$ and 
${\bf s}$.

The null states (corresponding to null fields) are defined as,
\be
{\cal W}_m^n|\chi_N>=0 ~~~\forall m>0, ~n\leq k, 
~~~ L_0|\chi_N>=(\Delta+N)|\chi>
\ee
By standard argument, they have null norm with any state in the Verma module.
The existence of such states depends on the choice of parameter $c_{\phi}$ and 
$\Delta_{\phi}$.

If one defines ${\bf\lambda}_{{\bf r},{\bf s}}$ as it follows
\bea
{\bf\lambda}_{{\bf r},{\bf s}}= \alpha_{+}{\bf r}+\alpha_{-}{\bf s}+
\alpha_0 {\bf\rho}
\eea
where $\alpha_{+}+\alpha_{-}=\alpha_{0}$ and $\alpha_{+}\alpha_{-}=-1$, 
the vectors $|\chi_{{\bf r},{\bf s}}>$ are singular with conformal weights
\be
\Delta_{\varphi}(\chi_{{\bf r},{\bf s}})=
\Delta_{\varphi}({\bf\lambda}_{{\bf r},{\bf s}})+{\bf r}\cdot{\bf s}
\ee
Null states can be obtained by using screening currents 
$:e^{i\alpha_{\pm}\alpha^i\varphi (z)}:$ of conformal dimensions one,   
where $\alpha^i$ are the simple roots of $sl(k)$ (notice that these roots are 
completely different from those of the $SU(2)_k$ algebra).

These vertex operators satisfy the property that the singular part in the OPE 
with $R_{k}(z)$ is a total derivative and then they can be inserted in any 
correlator without modifying the conformal properties (see \cite{FL,BS} for 
more detail). 

It is very easy to verify that the fields of eq:(\ref {eq: TWIST}) satisfy the 
following boundary conditions:
\be
\phi^{l}(\epsilon z^k) = \phi^{l+1}(z^k) ~~~~\forall ~~l=1,...,k
\ee 
which generate a cyclic permutation of the simple roots of the associated 
$sl(k)$.
 
This implies that a realization of ${\cal W}_{k}$ in terms of 
twisted bosons of the kind used by the authors in \cite{FKN} is needed.

This realization appears, for $\alpha_0=0$, in the infinite Grassmannian 
manifold approach to 2D quantum gravity and $k$-reduction procedure of KP 
hierarchy.  

The above twist conditions are diagonal for the following scalar fields:
\be
\phi_{l}(z^k) = \frac{1}{\sqrt k}\sum_{j=1}^{k}\epsilon^{j l}
\phi^{j}(z^k) ~~~~\forall ~~l=\{1,...,k-1\}
\ee
satisfying defined boundary conditions:
\be
\phi_{l}(\epsilon z^k) = \epsilon^{l}\phi_{l}(z^k) ~~~~\forall ~~l=\{1,...,k-1\}
\ee
with
\be
<\partial_{z}\phi_{l}(z^k)\partial_{w}\phi_{l'}(w^k)>=
\frac{1}{z w}
\left(\frac{z}{w}\right)^{l}\frac{\left((k-l)\frac{z}{w}+l\right)
}{\left(\left(\frac{z}{w}\right)^{k}-1\right)^{2}}\delta_{l+l',0(mod~k)}.
\ee

Moreover, the spin $n$ currents of ${\cal W}_{k}$ realized by using of these 
fields,  are single-valued under the above $Z_{k}$ conditions and can be 
written explicitly in terms of them expanding the OPE relations of 
parafermions eq.(\ref {eq: 38}) including also the regular terms (OPA).

The value of $\alpha_{0}^2=\frac{1}{(k+1)(k+2)}$ it is fixed to satisfy the  
central charge of $Z_{k}$ parafermions $c_{\psi}=2\frac{k-1}{k+2}$.

Lowest generators are, for example
\bea
{\cal W}^0(z^k)&=& 1 \\ 
{\cal W}^1(z^k)&=&\sum_{l=1}^{k}i\partial_{z}\phi^l(z^k) = 0 
\\ 
{\cal W}^2(z^k)&=&-\sum_{l=1}^{k}i\partial_{z}\phi^l(z^k)
i\partial_{z}\phi^{k-l}(z^k) 
+\alpha_{0}\sum_{l=1}^{k}(l-1)i\partial_{z}\phi^l(z^k) = 
T_{\psi}(z^k) \\ 
{\cal W}^3(z^k)&=&\sum_{l<l'<l''}^{k}i\partial_{z}\phi^l(z^k)
i\partial_{z}\phi^{l'}(z^k)i\partial_{z}\phi^{l''}(z^k) - 
\alpha_{0}\sum_{l=1}^{k}(l-1)i\partial_{z}\left (i\partial_{z}
\phi^l(z^k)i\partial_{z}\phi^{k-l}(z^k)\right ) \nn \\ 
&&- 
\alpha_{0}\sum_{l=1}^{k}(l-l'-1)i\partial_{z}\phi^l(z^k)
i\partial^2_{z}\phi^{k-l}(z^k) +
\frac{\alpha_0^2}{2} \sum_{l=1}^{k}(l-1)(l-2)i\partial^3_{z}\phi^l(z^k) 
\eea

It is possible to give an interesting interpretation of this realization for 
${\cal W}_{k}$ in terms of transverse projection of the covariant realization, 
following the arguments of sec.(2).

In fact, one can attack a light-like vertex $U^{mK^{+}}(z)$ to any pure 
transverse operator without modifying the OPE structure.

Therefore, I can give a realization of $W_{1+\infty}$ algebra (see 
\cite{AFMO} for more detail) in terms of the well known free boson 
realization \cite{FKN} for transverse part tensored with the light-like 
vertex:
\be
W^{n}_{mK^+}(z) = W^{n}(z)U^{mK^{+}}(z) ~~~\forall ~~m\in Z~~and ~~n\in Z_{+}
\ee
with $W^{n}(z)=:\prod_{i=1}^{k}Q^{(1)}(z):$.

By using the vacuum level $k$ projection of sec.(2), I obtain naturally a 
reduction to the transverse algebra.

Notice that, while the $X(z^k)$ component is unconstrained and thus gives a 
realization of the linear $W_{1+\infty}$ associated to the enveloping algebra 
of the abelian $\widehat{U}(1)$ Cartan sub-algebra, to make a complete 
identification between covariant and transverse realizations one must give the 
additional constraints:
\be
W^{n}_{mK^+}(z) |0>_{F^{k}} = 0  ~~~\forall ~~m\in Z~~and ~~n\in Z_{+} 
\label {eq: WN}
\ee
where $|0>_{F^{k}}$ is the vacuum state for Fock space of eq.(\ref {eq: 
FOCK}), implementing the request of orthogonality between ${\cal W}_{k}$ and 
the untwisted Cartan sub-algebra at the level $k$.

These are just the requests of sec.(5) of ref.\cite{FKN} to give a 
${\cal W}_{k}$ reduction for $\tau$ function of the KP hierarchy (in the 
$\alpha_{0}\rightarrow 0$ limit).

This reduction can be done in a formally well defined way by means of a 
standard procedure \cite{STT} using the operator $P_{X,k}$ to extract the 
parafermionic sub-algebra ${\cal W}_{k}$ from the full $W_{1+\infty}$.

The explicit form for this projector is given by:
\be
P_{X,k} = :\prod_{n>0}exp\left( \frac{a_{-n k}a_{n k}}{n k}\right):
\ee
with:
\be
P_{X,k}^{2} = 1, ~~~ P_{X,k} = P_{X,k}^{\dag} = P_{X,k}^{-1}
\ee

In the following, all the fields should be considered as obtained 
by means of this projector.

Notice, that this factorization is also verified for a special class of 
representations, called quasi-finite, of $W_{1+\infty}$ with a central charge 
$c=k$ \cite{AFMO}.

\subsection{Explicit realization} 

Higher spin currents introduced in this section can be extracted from the 
OPA of the currents constructed for the affine algebra to the $k$. 

Two aspect must be considered in this approach. The currents ${\cal W}_k$ 
should be considered on the $k$-sheeted plane, thus a Schwarzian terms appear 
in the definitions due to the different normal ordering associated to the 
transformation $z\rightarrow z^k$.

The second aspect is that in the case $\alpha_0\neq 0$ the fields $\phi^l(z^k)$ 
transforms in a non standard way for a logarithmic contribution:
\be
\phi^l(z^k) \rightarrow \phi^l(f(z^k)) + i \alpha_0 l ln f'(z^k)
\ee
so their derivatives become anomalous and cannot be extract straightforth from 
the OPE of vertex operators where they appear as total derivatives \cite{DF}.

Moreover, the currents in the KM algebra definitions are all single-valued so 
the anomalous terms are not evident in their OPE.
                                                                          
A formally well defined way to do this should be the introduction of $b$,$c$ 
ghost system and performing a BRST quantization.

In this paper I prefer to follow a more simple way that it seems to be more 
transparent and it gives a direct evidence of the existence of equivalence 
between covariant and transverse realizations that it is the aim of this paper.

Moreover, the effect of this anomaly is well evident in the the anomalous 
weight $\alpha\cdot\beta (1-1/k)$ appearing in eqs:(\ref{eq: 31},\ref{eq: PK}).
In the $SU(2)_k$ generators this term is always combined to the contribution of 
${\cal U}^{\alpha}(z^k)$ fields giving rise to the anomaly cancellation.

Therefore, I extract the generators form the OPA of the realization of 
$SU(2)_k$ algebra that appears only to the lowest order in $\alpha_0$ (zero 
order), by means of the identification with the operators defined in the 
eq.(\ref {eq: QM}).

For instance, using the definition of normal ordering:
\bea
&&\nor{{\cal U}^{-\alpha}(z^k)\psi^{-\alpha}_{k-\lambda}(z^k)
{\cal U}^{\alpha}(\xi^k)\psi^{\alpha}_{\lambda}(\xi^k)} + 
\nor{{\cal U}^{\alpha}(z^k)\psi^{\alpha}_{k-\lambda}(z^k)
{\cal U}^{-\alpha}(\xi^k)\psi^{-\alpha}_{\lambda}(\xi^k)} =  \\ 
&&\sum_{l,l'=1}^{k} \epsilon^{(l-l')\lambda}\oint_{|z|>|\xi|}
\frac{dz}{2 \pi i}
\frac{:{\cal U}^{-\alpha}(z^k){\cal U}^{\alpha}(\xi^k)
\psi^{-\alpha}(\epsilon^{l}z^k)\psi^{\alpha}(\epsilon^{l'}\xi^k):}{
(z-\xi)(\epsilon^{l}z-\epsilon^{l'}\xi)^2} + (\alpha \rightarrow -\alpha) 
\nn \\ 
&&= -\frac{1}{2}:\alpha\cdot H(z^{k})\alpha\cdot H(z^{k}): + \frac{1}{2}
z^{1-1/k}:\psi^{\alpha}(z^k)\partial_{z}^2z^{1-1/k}\psi^{-\alpha}(z^k):+c.c.
\eea

the equivalence
\be
\sum_{\alpha\in\Lambda_{+}}:\alpha\cdot H(z^{k})\alpha\cdot H(z^{k}):=
2 h^{\vee} \sum_{i=1}^{d} :H^{i}(z^{k})H^{i}(z^{k}):
\ee
(where $\Lambda_{+}$ is the positive roots lattice, $h^{\vee}$ is the dual 
Coxeter number $d$ is the rank of algebra) and the definition of Sugawara 
operators
\be
L_{m} = \frac{1}{2(k+h^{\vee})}\sum_{n\in Z}
\left (\sum_{i=1}^{d} :H^{i}_{n}H^{i}_{m-n}: + \frac{1}{2}
\sum_{\alpha\in\Lambda_{+}}
\nor{A^{\alpha}_{n} A^{-\alpha}_{m-n}} 
+\nor{A^{-\alpha}_{n} A^{\alpha}_{m-n}}\right), \label {eq: SUG}
\ee
where normal ordering it is defined as follows  
\bea
\nor{A^{\alpha}_{n} A^{-\alpha}_{m}} &=& 
A^{\alpha}_{n} A^{-\alpha}_{m} ~~~for~n\leq m \nn \\
\nor{A^{\alpha}_{n} A^{-\alpha}_{m}} &=& 
\frac{1}{2} \left (A^{\alpha}_{n} A^{-\alpha}_{m}+
A^{-\alpha}_{m} A^{\alpha}_{n} \right )~~~for~n= m\nn \\ 
\nor{A^{\alpha}_{n} A^{-\alpha}_{m}} &=& 
A^{-\alpha}_{m} A^{\alpha}_{n} ~~~for~n>m ~,
\eea

it is possible to show that the Sugawara expression corresponds to the 
projection on the $k$-sheeted plane of the usual free bosons construction 
(in this expression I specialize the realization to $SU(2)$ where the rank is 
one, the positive roots are simply $\alpha$ and $h^{\vee}=2$) corresponding to 
the following stress-tensor:
\bea
-\frac{1}{2} :i\partial_{z}X(z^k)i\partial_{z}X(z^k) -\frac{1}{2(k+2)}
\left (\sum_{l=1}^{k} 
:i\partial_{z}\phi_{l}(z^k)i\partial_{z}\phi_{k-l}(z^k): 
+i\alpha_0 \sum_{l=1}^{k}(l-1)\partial_{z}^2\phi_{l}(z^k)\right )
\nn \label {eq: WNP}
\eea
where I identify the second term in the r.h.s. of the equation with the 
${\cal W}^{2}_{m}$ operator.

While for $X(z^{k})$ term the central charge is always one, as the 
$\phi(z^{k})$ fields give contribution only by means of $A^{\alpha}_{n}$ 
operators, they cannot be rescaled without breaking the correct commutation 
relations that seem to be necessary to recover the charge corresponding to the 
parafermions $c_{\psi}=\frac{h^{\vee}(k-1)}{k+h^{\vee}}=\frac{2(k-1)}{k+2} $. 

To obtain the standard expression of eq:(\ref {eq: QM}) it is necessary to make 
use of the full properties of conformal theory; in fact, null states existing 
at any level $N={\bf r}\cdot{\bf s}$ in completely degenerate representations 
of enhanced algebra ${\cal W}_k$  can be useful to change the 
form of the generators. 

For instance, in the $k=2$ case, where only the conformal algebra exists, the  
lowest states are:
\bea
|\chi_{11}>&=&L_{-1}|h_{11}> \nn \\
|\chi_{21}>&=&(L_{-2}-\frac{3}{4}L^2_{-1})|h_{21}> \nn \\
|\chi_{12}>&=&(L_{-2}-\frac{4}{3}L^2_{-1})|h_{12}> \nn \\
\eea

One can use the freedom to add any superposition of null fields to OPE to put 
the Virasoro generators in the usual form.
In the above example the state is simply 
\be
\frac{8}{3}|\chi_{21}>-\frac{3}{2}|\chi_{12}>= \frac{1}{2}L_{-2}
(|h_{21}>-|h_{12}>),
\ee
thus the Sugawara stress-tensor and the Miura expression become equivalent 
modulo conformal null fields.

An interesting consequence of this decomposition is that the well known 
equivalence between Sugawara and Virasoro construction of conformal algebra 
for $k=1$ can be extended to any $k$ by means of this reduction.

At this point it is possible to give an example of calculus of higher order 
contributions from OPA of $SU(2)_k$ currents.

Spin three generator comes from the following Sugawara like term:
\bea
&&\frac{1}{2(k^{3/2}+2^{3/2})}\oint_{|z|>|\xi|}\frac{dz}{2 \pi i}
\left [\frac{:H(z)H(\xi):}{(z-\xi)}  \right.\\ 
&& \left. + \sum_{l,l'=1}^{k} 
\frac{:{\cal U}^{-\alpha}(z^k){\cal U}^{\alpha}(\xi^k)
\psi^{-\alpha}(\epsilon^{l}z^k)\psi^{\alpha}(\epsilon^{l'}\xi^k):}{
(z-\xi)^2(\epsilon^{l}z-\epsilon^{l'}\xi)^2} 
- (\alpha \rightarrow -\alpha ) \right ] \nn \\ 
&&= -\frac{1}{6}:i\partial_{z}X(z^{k})i\partial_{z}X(z^{k})
i\partial_{z}X(z^{k}): 
+\frac{(\sqrt{k}+\sqrt{2})}{6(k^{3/2}+2^{3/2})} i\partial_{z}^{3}X(z^{k}) 
\nn \\ 
&&-
\frac{2^{3/2}}{6\sqrt{k}(k^{3/2}+2^{3/2})}\sum_{l+l'+l''=k}^{k} 
:i\partial_{z}\phi_{l}(z^k)i\partial_{z}\phi_{l'}(z^k)
i\partial_{z}\phi_{l''}(z^k): \nn \\ 
&&+ \frac{2^{3/2}}{6(k^{3/2}+2^{3/2})}i\partial_{z}X(z^{k})\sum_{l=1}^{k} 
:i\partial_{z}\phi_{l}(z^k)i\partial_{z}\phi_{k-l}(z^k): + {\cal O}(\alpha_0)
\eea
where the normalization is chosen to put the $X(z^k)$ currents in the 
standard basis.

This expression has a clear interpretation in terms of ${\cal W}_{k}$ 
generators.
The first two terms come from the $\widehat{U}(1)$ while the third must be 
considered as $\alpha_0$ lowest order of ${\cal W}^3(z^k)$ and finally the 
last is the coupling between $X(z^k)$ current and the stress-tensor of 
parafermions ${\cal W}^2(z^k)$.

To put the above expression of ${\cal W}^3(z^k)$ in the standard form, it is 
possible to make use of null fields as it was done for the stress-tensor where 
now a superposition of the level $rs=3$ must be used.

Performing this procedure up to the level $n=k$ one recovers the Miura 
realization of eq:(\ref {eq: QM}) in terms of the above parafermion fields in 
completely degenerate representations.
Unfortunately I do not know of a general explicit form for null states in the 
arbitrary $k$ case.

Moreover, the sum over the normal ordered product of non-abelian currents is 
always equivalent to a sum over the abelian algebra of Cartan sub-algebra of 
the realization of level one, with the additional constraint of taking only 
the combinations that are singlet for the discrete algebra $Z_{k}$.

The general structure appears as the product of $\widehat{U}(1)$ enveloping 
algebra for $X(z^k)$ and a ${\cal W}_k$ for parafermions.

For large value of $k$, $c_{\phi}\rightarrow 2$ and the underlying 
${\cal W}_k$ algebra linearizes in the $k\rightarrow \infty$ limit giving rise 
to the well known $Z_{\infty}$ parafermionic realization of $W_{\infty}$ 
\cite{BK}.

Naturally, all these arguments also hold for the dual realization obtained by 
using of the multi-value fields where more complications arise for the not 
analyticity of the involved functions.
Nevertheless, the isomorphism of eq:(\ref{eq: 32}) assures the exact 
cancellation of all these contributions.   

\bigskip

\section{Conclusions and Further Remarks}

In this paper I discuss some interesting properties of the extension to 
${\cal W}_{k}$ algebra of the previous studied equivalence between covariant 
and light-cone gauge realization of Kac-Moody algebras.

Many interesting observations can be done for these models. At the first, it 
is evident that the previous analysis can be extended to the higher spin 
generators that are obtained by means of $k$-projection of a product of $d$ 
Fubini fields $Q(z)$.

For algebras of rank larger then one there are new independent terms coming 
from these generators also at level $k=1$.

For instance, to the algebra $SU(3)$ one can associate a ${\cal W}^{3}(z)$ 
in terms of the $X(z^k)$ fields using the third order invariant Casimir 
tensor, the ${\cal W}_{c}[\hat{g}/g,1]$ algebra obtained in this way is 
denoted as Casimir algebra at level one \cite{BBSS}.
When $k>1$ one needs to add also contributions coming from the higher terms in 
the operator product algebra of parafermions.

Therefore, in the above mentioned example of $SU(3)$ to the level $k$, there 
are two commuting ${\cal W}^{3}(z)$ fields, one is obtained from the $X(z^k)$ 
and the second one depends on the $\phi(z^k)$ fields.

It should also be noted that, the full generator must be a superposition of 
these terms which, for all the case of spin $n>2$, contain also mixed operators 
depending on both the kind of fields.

For an arbitrary level $k>1$ it should be reasonable to identify this algebra 
with the Casimir algebra ${\cal W}_{c}[\hat{g}/g,k]$, but the identification 
with the well known infinitely fields generated algebra arising in this case 
is quite obscure.

Another interesting aspect that arises from this realization is the extension 
of the quantum equivalence existing in the $k=1$ case between Sugawara and 
Virasoro construction of conformal algebra to all the level $k$ of sec.(3) and 
to all higher spin currents.

One notes that, at least for the unitary series, the restriction to the Cartan 
sub-algebra at level one of Casimir tensors used in the construction of 
${\cal W}_{c}[\hat{g}/g,1]$ algebras is enough to give ${\cal W}_{k}$ 
operators to any $k$ when a proper projection is taken (as it is explained in 
sec.(3)).

Therefore, one needs to consider a new class of traceless symmetric tensors 
$T^{\{a,b,...,c\}}_{k}(z)$  defined on the compactification space of a string 
theory.
Indices $\{a,b,...,c\}$ run on $R^{d}$ (where $d$ is the rank of the 
algebra), while the variable $z$ indicates that the tensor is not a constant 
for two-dimensional world-sheet; moreover, only for level one the Casimir 
tensors should be chosen to be constant, as it follows from the construction 
and from the observation that they must be invariant tensors only for the 
finite sub-algebra and not for the full affine algebra.

The lowest degree tensor in this class is just the usual metric tensor on the 
Casimir sub-algebra $T^{\{i,j\}}_{1}(z)=g^{ij}$  which for $SU(2)$ is also the 
unique possible, but for $k>1$ it should become a local field $g^{ij}(z)$ 
obtained by means of a proper projection. 

This implies that the natural two dimensional geometry for these models should 
not be the usual Riemann sphere but rather a branch covering the Riemann sphere 
that can be interpreted as affine algebraic curve (for $SU(2)_k$ 
this is just a $Z_k$ symmetric algebraic curve \cite{FSK}). 
These arguments should give a direct application of the present realization to 
the recently analyzed twisted-WZW models on elliptic curves \cite{KT} and 
related generalizations.   

Traceless symmetric tensors $T^{\{a,b,...,c\}}_{k}(z)$ can be factorized in the 
$T_{X}$ and $T_{\phi}$ components (for algebras with rank larger than one, 
generally, also tensors $T_{X,\phi}$ coupling $X$ and $\phi$ fields exist).

It should also be interesting to study in detail the representation theory 
for these cosets, to understand more deeply the structure and the connections 
with Lorentzian algebras which, recently, have been recognized to be related to 
non-perturbative effects in string theory \cite{DVV,HM}.

\bigskip

{\bf Acknowledgments} - The author is indebted to A. Sciarrino for useful comments and 
the reading of the manuscript.

\end{document}